\documentclass[12pt]{article}
\usepackage{amsmath,amssymb}
\usepackage{cite}
\usepackage{graphicx,color}
\usepackage{setspace}
\usepackage{nicefrac}
\usepackage[nottoc]{tocbibind} % add references to TOC

\usepackage[linktoc=all,hidelinks]{hyperref}

\usepackage{indentfirst}
\usepackage[lf]{Baskervaldx} % lining figures
\usepackage[bigdelims,vvarbb]{newtxmath} % math italic letters from Nimbus Roman
\usepackage[cal=boondoxo]{mathalfa} % mathcal from STIX, unslanted a bit

\usepackage{geometry}
\def\p{\partial}
\def\m{\mu}
\def\n{\nu}

\def\a{\alpha}
\def\b{\beta}
\def\d{\delta}

\def\e{\eta}

\def\ep{\epsilon}
\def\ve{\varepsilon}
\def\f{\phi}
\def\F{\Phi}
\def\k{\kappa}

\def\L{\Lambda}
\def\pr{\prime}

\def\r{\rho}

\def\t{\theta}

\def\vf{\varphi}
\def\z{\zeta}
\def\nn{\nonumber}
\def\sq{\sqrt}
\def\sqdet{\sq{-g}}
\def\sqeta{\sq{-\eta}}

\def\goesto{\rightarrow}

\def\cL{\mathcal{L}}

\def\wed{\wedge}
\def\imp{\implies}

\newcommand{\RHS}{\text{(RHS)}}
\newcommand{\eff}{\text{eff}}
\newcommand{\cons}{\text{constant}}
\newcommand{\dual}{\text{dual}}
\newcommand{\New}{\text{N}}
\newcommand{\ma}{\text{max}}
\newcommand{\mi}{\text{min}}

\newcommand{\link}[1]{[\href{http://arxiv.org/abs/#1}{{\tt arXiv:#1}}]}
\newcommand{\linkth}[1]{[\href{http://arxiv.org/abs/hep-th/#1}{{\tt arXiv/hep-th:#1}}]}

\newcommand{\doi}[1]{[\href{http://doi.org/#1}{{\tt doi:#1}}]}
\newcommand{\mail}[1]{\href{mailto:#1}{{\tt #1}}}

\newcommand*\dif{\mathop{}\!\mathrm{d}}

\numberwithin{equation}{section}

\usepackage{titletoc}
\dottedcontents{section}[1.2em]{}{1em}{0.5pc}
\dottedcontents{subsection}[4em]{}{2em}{1pc}

\begin{document}

\begin{titlepage}       %\vspace{5pt} \hfill 

%\vspace{5mm}

\begin{center}
	\setstretch{2}
	{\LARGE \bf Kerr-Schild Double Copy of the Coulomb Solution in Three Dimensions}
\end{center}

\begin{center}
\vspace{10pt}

{G{\"o}khan Alka\c{c}$\,{}^{a}$, Mehmet Kemal G\"{u}m\"{u}\c{s}$\,{}^{b}$ and Mehmet Ali Olpak$\,{}^{c}$}
\\[4mm]

{\small 
{\it ${}^a$Physics Engineering Department, Faculty of Engineering,\\ Hacettepe University, 06800, Ankara, Turkey}\\[2mm]

{\it ${}^b$Department of Physics, Faculty of Arts and Sciences,\\
	Middle East Technical University, 06800, Ankara, Turkey}\\[2mm]

{\it ${}^c$Department of Electrical and Electronics Engineering, Faculty of Engineering,\\ University of Turkish Aeronautical Association, 06790, Ankara, Turkey}\\[2mm]
e-mail: {\mail{gokhanalkac@hacettepe.edu.tr}, \mail{kemal.gumus@metu.edu.tr}, \mail{maolpak@thk.edu.tr}}
}
\vspace{5pt}
\end{center}

\centerline{{\Large \bf{Abstract}}}
\vspace*{5mm}
While the Kerr-Schild double copy of the Coulomb solution in dimensions higher than three is the Schwarzschild black hole, it is known that it should be a nonvacuum solution in three dimensions. We show that the static black hole solution of Einstein-Maxwell theory (with one ghost sign in the action) is the double copy with the correct Newtonian limit, which provides an improvement over the previous construction with a free scalar field that does not vanish at infinity. By considering a negative cosmological constant, we also study the charged Ba\~nados-Teitelboim-Zanelli black hole and find that the single copy gauge field is the Coulomb solution modified by a term which describes an electric field linearly increasing with the radial coordinate, which is the usual behavior of the Schwarzschild-AdS black hole in higher dimensions when written around a flat background metric.

%\vspace{10pt}

	\tableofcontents

\end{titlepage}

%{\bf \large	
\section{Introduction}
After its discovery in the context of scattering amplitudes \cite{Bern:2010ue,Bern:2010yg}, the double copy idea has found applications in the classical domain as perturbative and exact relations between gravity and gauge theories. While perturbative spacetimes can be obtained by squaring the numerator of some diagrams in Yang-Mills theory \cite{Monteiro:2014cda,Luna:2016hge,Goldberger:2016iau,Goldberger:2017frp,Goldberger:2017vcg,Goldberger:2017ogt,Shen:2018ebu,Carrillo-Gonzalez:2018pjk,Plefka:2018dpa,Plefka:2019hmz,Goldberger:2019xef,PV:2019uuv,Anastasiou:2014qba,Borsten:2015pla,Anastasiou:2016csv,Cardoso:2016ngt,Borsten:2017jpt,Anastasiou:2017taf,Anastasiou:2018rdx,LopesCardoso:2018xes,Luna:2020adi,Borsten:2020xbt,Borsten:2020zgj,Luna:2017dtq, Kosower:2018adc, Maybee:2019jus, Bautista:2019evw, Bautista:2019tdr, Cheung:2018wkq, Bern:2019crd, Bern:2019nnu, Bern:2020buy,Kosmopoulos:2021zoq, Kalin:2019rwq, Kalin:2020mvi, Almeida:2020mrg, Godazgar:2020zbv, Chacon:2020fmr}, it is possible to establish a map between exact solutions that take a much simpler form, where certain classical solutions of general relativity (GR) are in correspondence with solutions of Maxwell's theory \cite{Luna:2015paa,Bahjat-Abbas:2017htu,Carrillo-Gonzalez:2017iyj,Alkac:2021bav,Luna:2016due,Berman:2018hwd,Bah:2019sda,Banerjee:2019saj,Ilderton:2018lsf,Lee:2018gxc,Cho:2019ype,Lescano:2020nve,Lescano:2021ooe,Kim:2019jwm,Keeler:2020rcv,Huang:2019cja,Alawadhi:2019urr,Luna:2018dpt,Alawadhi:2020jrv,Elor:2020nqe,Berman:2020xvs,Easson:2020esh,CarrilloGonzalez:2019gof,Gumus:2020hbb}. Among different versions of the classical double copy that have been discovered so far, the Kerr-Schild (KS) double copy \cite{Monteiro:2014cda} has the advantage of admitting a formulation in arbitrary dimensions. For metrics of the KS form \cite{Kerr}, the Ricci tensor with mixed indices is linear in the perturbation for a flat \cite{Gurses:1975vu} and curved \cite{taub,Xant1,Xant2} background metric. Due to this crucial property, a map to solutions of Maxwell's theory defined on the background spacetime, which has a linear nature, can be established. Alternatively, by writing the tensorial equations of GR and Maxwell's theory in $d=4$ in terms of $2$-component spinors, one can construct a Weyl spinor characterizing the spacetime from two electromagnetic field strength spinors up to a scalar field, leading to the Weyl double copy \cite{Luna:2018dpt}. Also,  a class of solutions of GR and self-dual solutions of Maxwell's theory can be related through  Newman-Penrose formalism \cite{Newman:1961qr}, giving rise to the Newman-Penrose map \cite{Elor:2020nqe}. Although the last two are limited to $d=4$ in their present formulations, the fact that they were shown to originate from twistor theory \cite{White:2020sfn,Chacon:2021wbr,Farnsworth:2021wvs} is a sign of a much richer structure than previously thought, which already found a practical utility by providing a way to fix the scalar field in the Weyl double copy. 

Having a new theoretical tool to investigate the relation between two seemingly different theories, it is important to consider situations where it might break in order to understand limitations if there are any. A possible route in this direction is to work in three dimensions (3d), for which the KS formulation is available, and  this was pursued in \cite{CarrilloGonzalez:2019gof,Gumus:2020hbb}. Due to the lack of degrees of freedom and a Newtonian limit in 3d GR,  it is not obvious, at first sight, how the procedure works.\footnote{See \cite{Moynihan:2020ejh,Burger:2021wss} for the study of 3d amplitudes where a degree of freedom is introduced in the gravity side by adding a Chern-Simons term in the action. As a result, one obtains the amplitudes in topologically massive gravity as the double copy of the amplitudes of topologically massive electrodynamics.} For the Coulomb solution, which is the simplest nontrivial solution in the gauge theory side, the gauge boson degrees of freedom is mapped to the graviton and the KS scalar characterizing the spacetime metric is directly linked to the Newtonian potential as given in \eqref{newt}. However, without matter coupling, this cannot be realized since the Newtonian potential vanishes identically. When the problem was tackled by coupling a free scalar field, a hairy black hole with the desired properties can be obtained if the scalar is a ghost, which is equivalent to coupling a spacelike fluid \cite{CarrilloGonzalez:2019gof}. In order to support the black hole solution, the scalar should be linear in the azimuthal angle, and therefore, does not vanish at infinity as suggested by the no-hair theorem. In the linearized theory, the ghost sign can be removed by a certain generalized gauge transformation; however, it is still somewhat unsatisfactory to not have a reasonable behavior of the matter field at infinity. When investigated further in \cite{Gumus:2020hbb}, it was proposed to take the Einstein-Hilbert term with a ghost sign, which does not introduce a dynamical ghost in the theory and removes the need for a generalized gauge transformation to get rid of the ghost in the linearized theory. 

In this paper, we aim to present an alternative for the matter coupling with a better behavior at infinity, which, as we will see, also provides a beautiful connection to a well-known solution of 3d black hole physics. In Sec. \ref{sec:sca}, we will make a review of the construction of \cite{CarrilloGonzalez:2019gof} by emphasizing the points that will be relevant to our later discussion. In Sec. \ref{sec:dual}, considering the on-shell duality of a scalar and a gauge vector together with the KS ansatz for the metric, we will find that the same type of solution can be obtained in Einstein-Maxwell theory. In Sec. \ref{sec:EM}, we will obtain the most general static solution of the KS form by introducing a cosmological constant. When the cosmological constant is zero, we will show that a charged black hole solution with the correct Newtonian limit, which also gives rise to the Coulomb solution as its single copy, can be obtained when a ghost sign is used in the action. The electric field corresponding to the gauge field in the gravity side vanishes at infinity, and therefore, provides the promised improvement. For a negative cosmological constant, the charged Ba\~nados-Teitelboim-Zanelli black hole \cite{Banados:1992wn} follows from the most general solution without taking any ghost sign in the action. We end this section by studying the gauge theory single copy of the solution. Finally, we present our conclusions in Sec. \ref{sec:conc}.

\section{The Coulomb Solution from the Free Scalar}\label{sec:sca}
\subsection{Solution with the correct Newtonian potential}
In \cite{CarrilloGonzalez:2019gof}, the Coulomb solution was obtained as the single copy of the static black hole solution of GR coupled to a free scalar with the following action:
\begin{equation}
S=\int \dif^{3}x\sqrt{-g}\left[\frac{\z_{1}}{\kappa^{2}}R-\frac{\z_2}{2}\left(\p \vf\right)^2\right], \qquad \k^2=8\pi G\label{actsca}
\end{equation}
where $\z_i=\pm 1$ ($i=1,2$) control the sign of the kinetic terms and take a negative value for a ghost graviton or a dilaton \cite{Gumus:2020hbb}. The  field equations which follow from the action \eqref{actsca} are
\begin{align}
R_{\m\n}&=\z\frac{\k^2}{2} \p_\m \vf \p_\n \vf,\label{eqgravsc}\\
\p_\m(\sqdet\, g^{\m\n}\p_\n \vf)&=0,\label{eqsc}
\end{align}
where $\z=\z_{1} \z_2$. Let us consider the following static KS metric around the Minkowski space
\begin{equation}
g_{\m \n}=\eta_{\mu \nu}+\f(r)\, k_\m k_\n\,,\label{KS}
\end{equation}
where the vector $k_\mu$ is null and geodesic with respect to both the background metric $\eta_{\mu \nu}$ and the full metric $g_{\mu \nu}$ (see chap. 32 of \cite{Stephani:2003tm} for a detailed discussion of important properties). Writing the background line element in polar coordinates
\begin{equation}
\e_{\m\n}\dif x^\m \dif x^\n=-\dif t^2+\dif r^2+r^2\dif \t^2,\label{minkpol}
\end{equation}
the vector $k_\m$ can be written as follows
\begin{equation}
	-k_\m \dif x^\m=\dif t + \dif r,\label{kdef}
\end{equation}
and the line element in the KS coordinates becomes
\begin{align}
	\dif s^2 &= \e_{\m\n} \dif x^\m \dif x^\n + \f(r) (k_\m \dif x^\m)^2\nn\\
	&= -\left[1-\f(r)\right] \dif t^2 + \left[1+\f(r)\right]  \dif r^2 +2\f(r) \dif t \dif r+r^2\dif \t^2.
\end{align}
If one assumes\footnote{In \cite{CarrilloGonzalez:2019gof}, the authors directly use the KS scalar yielding the correct Newtonian potential and conclude that the solution should be sourced by a free scalar. Here, we give the derivation in a way that will be useful in our later discussion.} 
\begin{equation}
	\p_\m \vf = (0,0,c), \qquad \qquad c=\text{constant},\label{scasol}
\end{equation}
the equation for the scalar field \eqref{eqsc} is satisfied independent of the KS scalar $\f$ as follows:
\begin{equation}
	\p_\m(\sqdet\, g^{\m\n}\p_\n \vf)=\p_\m(\sqeta\, \e^{\m\n}\p_\n \vf)=0,\label{scaeq}
\end{equation}
where we have used $\det g = \det \e$, $g^{\t\t} = \e^{\t\t}$ and $g^{t \t } = g^{r \t}=0$.

With this at hand, one can now check the gravity equations \eqref{eqgravsc}. The independent nonzero components of the Ricci tensor read
\begin{align}
	R_{tt} &= \frac{\left[\f(r)-1\right]\left[r\,\f^{\pr \pr}(r)+\f^\pr(r)\right]}{2r},\nn\\
	R_{tr} &=\frac{\f(r)\left[r\,\f^{\pr \pr}(r)+\f^\pr(r)\right]}{2r},\nn\\
	R_{rr} &=\frac{\left[\f(r)+1\right]\left[r\,\f^{\pr \pr}(r)+\f^\pr(r)\right]}{2r}, \nn\\
	R_{\t\t} &= r\, \f^\pr(r),
\end{align}
and the only nonzero component of the right-hand-side of the equations is
\begin{equation}
	\RHS_{\t\t} = 4\pi \z G c^2, 
\end{equation}
From the $\t\t$ component, one finds the solution for the KS scalar as
\begin{equation}
	\f = b+4\pi \z G c^2 \log(r),\qquad b = \text{constant},
\end{equation}
which also satisfies the remaining components. The constant $c$ and the parameter $\z$ can be fixed by considering the Newtonian limit. The Newtonian potential is given by
\begin{equation}
	\F = -\frac{1}{2}(1+g_{00})=-\frac{\f}{2},\label{newt}
\end{equation}
and in order to mimic the Newtonian gravity 
\begin{equation}
	\vec{g} =- \vec{\nabla} \F = -\frac{GM}{r}\hat{r},\label{g}
\end{equation}
the KS scalar should be in the following form:
\begin{equation}
	\f = -2 G M \log(r) + \text{constant}.\label{KSnewt}
\end{equation}
Therefore, we need to fix the parameters as follows:
\begin{equation}
	c = \sq{\frac{M}{2\pi}}, \qquad \qquad \z = -1,\label{scaparam}
\end{equation}
with which, the KS scalar becomes
\begin{equation}
	\f = -2GM \log(r)+b.
\end{equation}
Since $\z=-1$, one should choose the ``wrong sign'' for one of the kinetic terms in the action \eqref{actsca}. While the scalar was chosen to be a ghost in \cite{CarrilloGonzalez:2019gof}, introducing the Einstein-Hilbert (EH) term with the negative sign has the advantage that it does not propagate any physical degree of freedom \cite{Gumus:2020hbb}.

In order to fix the integration	 constant $b$, we write the metric \eqref{KS} in the Boyer-Lindsquit (BL) coordinates by the following coordinate transformation:
\begin{equation}
\dif t \rightarrow \dif t +\frac{\f(r)}{1-\f(r)}\, \dif r,\label{trans}
\end{equation}
which leads to the line element
\begin{equation}
\dif s^2 = -f(r) \dif t^2 + \frac{\dif r^2}{f(r)}+r^2 \dif \t^2, \qquad f(r) = 1 - \f(r),\label{BL}
\end{equation}
for a generic KS scalar. In our case, the metric function becomes
\begin{equation}
	f(r)=1-b+2GM \log(r).
\end{equation}
In order to recover the Minkowski spacetime when the black hole mass vanishes ($M=0$), one should take $b=0$. Therefore, the KS scalar and the metric function are given by
\begin{align}
	\f(r)&=-2GM \log(r),\\
	f(r)&=1+2GM\log(r).\label{fsca}
\end{align}
We refer the reader to \cite{CarrilloGonzalez:2019gof} for an analysis of the motion of massive particles where the authors show that stable orbits exist for a certain range of parameters. In Sec. \ref{sec:EM}, we will show that it is true for the solution of Einstein-Maxwell theory.
\subsection{Gauge theory single copy}
The gauge theory single copy for a generic matter coupling can be obtained by considering the trace-reversed gravity equations 
\begin{equation}
R^{\m}_{\ \n}=\frac{\k{^2}}{2}\left[T^{\m}_{\ \n}-\d^{\m}_{\ \n}T\right],\qquad \qquad T = T^{\m}_{\ \m},\label{reversed}
\end{equation}
where $T_{\m\n}$ is the energy-momentum tensor. For a KS metric \eqref{KS}, the Ricci tensor with mixed indices reads
\begin{equation}
R^{\m}_{\ \n}=\frac{1}{2}\left[\p^\a \p^\m (\f\,k_\n k_\a) + \p^\a  \p_\n (\f\,k^\m k_\a) -\p^\a\p_\a(\f\,k^\m k_\n)\right],
\end{equation}
which is linear in the perturbation. If  $k^0=+1$ and one identifies $A_\m = \f\, k_\m$, the $\m0$ component can be written as
\begin{equation}
R^{\m}_{\ 0} = \frac{1}{2}\p_\n F^{\n\m},\label{delF}
\end{equation}
where $F_{\m\n}=2\p_{[\m}A_{\n]}$ is the field strength tensor of the gauge field $A_\m$. Therefore, the $\m0$ component of the gravity equations can be mapped to Maxwell's equations as follows:
\begin{equation}
\p_\n F^{\n\m} = g J^\m,
\end{equation}
where the source is given by
\begin{equation}
J^\m=2\left[T^{\m}_{\ 0}-\d^{\m}_{\ 0} T\right],\label{Jdef}
\end{equation}
and the gauge coupling is obtained by $\frac{\k^2}{2}\goesto g$ \cite{Monteiro:2014cda}. 

Application of this procedure to our solution gives the single copy gauge field as\footnote{We write the solution with a different normalization than \cite{CarrilloGonzalez:2019gof,Gumus:2020hbb} to simplify the solution of Einstein-Maxwell theory that we will give in Sec. \ref{sec:EM}. The Maxwell action is taken as \eqref{freevec} and we formulate the scalar-vector duality in Sec. \ref{sec:dual} accordingly.} 
\begin{equation}
	A_\m\dif x^\m =\f k_\m\dif x^\m=  Q \log r\,(\dif t + \dif r),\label{coulomb}
\end{equation}
 where we have made the replacement $2 G M \goesto Q$. This is just the Coulomb solution in a gauge where $A_\m A^\m = 0$, and the source is
 \begin{equation}
 	J^\m \p_{\m} = Q\, \d^2(\vec{r})\, \p_t,
 \end{equation}
 which corresponds to a charged particle in the flat spacetime.
 
Although we achieved the Coulomb solution as the single copy, the construction has the undesired feature that the scalar field in the gravity side is linear in the azimuthal angle. This is not unexpected due to the existence of the scalar hair (see Appendix of \cite{Gumus:2020hbb} for a detailed discussion of no-hair theorem for free scalar fields). One way to obtain a matter configuration which is well behaved at infinity is to consider the coupling of a gauge vector since it will yield a global charge, i.e., the electric charge, and therefore, produces no hair. In $d\geq4$, this is just the Reissner-Nordstr\"{o}m black hole solution of Einstein-Maxwell theory, whose metric can be also written in the KS form around Minkowski background.

\section{Scalar-Vector Duality and Its Consequences}\label{sec:dual}
In this section, we will present the duality between a free scalar and a gauge vector in three dimensions by following Sec. 7.8 of \cite{Freedman:2012zz}, and then, discuss its consequences for the classical double copy. In $d$-dimensions, the number of on-shell degrees of freedom of a $p$-form gauge field is $C(d-2,p) = \frac{(d-2)!}{p!(d-p-2)!}$. Due to the identity $C(d-2,p)=C(d-2,d-p-2)$, a $p$-form and a $(d-p-2)$-form in $d$-dimensions have the same number of degrees of freedom. In $d=3$, this implies that a scalar ($p=0$) and a vector gauge field ($p=1$) have the same number of degrees of freedom, which is one. Indeed, one can also show that the free field equations are equivalent and the solutions are in one-to-one correspondence. In order to see that, let us consider the following flat space action
\begin{equation}
S=\int \dif^{3}x\sqeta\left[\frac{1}{8 \pi}f_{\m\n}f^{\m\n}+\frac{1}{2\sqrt{2\pi}}\ep^{\m\n\r}f_{\m\n} \p_\r\vf \right],\label{dualact}
\end{equation}
where $\ep^{\m\n\r} = \frac{1}{\sqeta} {\ve^{\m\n\r}}$ is the Levi-Civita tensor and we take the Minkowski spacetime in polar coordinates \eqref{minkpol} for later convenience. The equation for $\vf$ gives
\begin{equation}
\ep^{\m\n\r}\p_{\m}f_{\n\r}=0,
\end{equation}
which implies $f_{\m\n}=2\p_{[\m}a_{\n]}$, i.e., $f_{\m\n}$ is the field strength tensor of a gauge field $a_\m$. Checking the equation for  $f_{\m\n}$ gives how it is related to the scalar $\vf$ as follows:
\begin{equation}
f_{\m\n}=-\sqrt{2\pi}\,\ep_{\m\n\r}\p^\r \vf \imp \p_\m \vf=\frac{1}{2\sqrt{2\pi}}\, \ep_{\m\n\r} f^{\n\r}.\label{dual}
\end{equation}
Inserting the expression for $f_{\m\n}$ into the action \eqref{dualact} yields the action for a free scalar 
\begin{equation}
S_{\text{scalar}}=\int \dif^{3}x\sqeta\, \left[\frac{1}{2}\left(\p\vf\right)^2\right],\label{freesca}
\end{equation}
with the field equation
\begin{equation}
	\p_\m(\sqeta\, \e^{\m\n}\p_\n \vf)=0.
\end{equation}
In the same way, one can also eliminate $\vf$ from the action \eqref{dualact} by using the expression in \eqref{dual}, which yields the action for a free gauge field
\begin{equation}
S_{\text{vector}}=\int \dif^{3}x\sqeta\,\left[-\frac{1}{8 \pi}f_{\m\n}f^{\m\n}\right],\label{freevec}
\end{equation}
where the field equation is given by
\begin{equation}
	\p_\n(\sqeta\, \e^{\n\a}\e^{\m\b}f_{\a\b})=0.
\end{equation}

We are now in a position to discuss the implications of the duality. As we have seen in  \eqref{scaeq}, the scalar field configuration  given in \eqref{scasol} is a solution when the spacetime is curved and endowed with the metric \eqref{KS}, or equivalently, flat and endowed with the Minkowski metric in polar coordinates \eqref{minkpol}. Our analysis shows that the actions for the free scalar \eqref{freesca} and the free vector \eqref{freevec} are equivalent and the solutions to free field equations (\ref{freesca},\ref{freevec}) are in one-to-one correspondence where the relation between the solutions is given in \eqref{dual}. For the solution of the scalar field given in \eqref{scasol}, this implies that,  for the corresponding vector solution, the only nonzero component of the field strength tensor is
\begin{equation}
	f_{tr} \propto \frac{1}{r},
\end{equation}
which is the electric part. Introducing the electric charge as the proportionality constant, the gauge field and the field strength tensor can be written as
\begin{align}
	a_\m \dif x^\m &= q \log(r) \dif t,\nn\\
	 \frac{1}{2}\,f_{\m\n}\, \dif x^\m \wed \dif x^\n &= \frac{q}{r}\, \dif r \wed \dif t.\label{elec}
\end{align}
Similar to the scalar case, this field configuration is also a solution when the spacetime is curved and endowed with the metric \eqref{KS} as follows:
\begin{equation}
\p_\n(\sqeta\, \e^{\n\a}\e^{\m\b}f_{\a\b})=\p_\n(\sqdet \, g^{\n\a}g^{\m\b}f_{\a\b})=0.
\end{equation}
As emphasized in \cite{Freedman:2012zz}, although the equivalence is true for the simplest kinetic actions, it is not guaranteed to hold in a more general setup. We have proven that the scalar solution in curved spacetime with a KS metric implies that the electric field configuration \eqref{elec} is also a solution in such a spacetime. However, one should keep in mind that this statement is independent of the KS scalar $\f(r)$. Therefore, when coupled to gravity, the matter equations will definitely be satisfied if the metric can be put in the KS form \eqref{KS}; however, the KS scalars in these two cases might differ. Indeed, one immediately sees that the static black hole solution obtained by coupling the vector field to gravity should be a charged black hole solution. In the next section, we will study the black hole solution with this new matter coupling.

\section{Einstein-Maxwell Theory in Three Dimensions}\label{sec:EM}

\subsection{The most general static solution of Kerr-Schild form}
Motivated by the results of the previous section, we consider Einstein-Maxwell theory with a cosmological constant described by the following action:
\begin{equation}
S=\int \dif^{3}x\sqrt{-g}\left[\frac{\z_1}{\k^2}\left(R-2\L\right)-\frac{\z_2}{8 \pi}f_{\m\n}f^{\m\n}\right], \qquad \k^2 = 8\pi G,\label{acvec}
\end{equation}
where, similar to the scalar case, $\z_i=\pm 1$ ($i=1,2$) control the sign of the kinetic terms and take a negative value for a ghost graviton or a vector. The field equations arising from the action \eqref{acvec} are given by
\begin{align}
R_{\m\n}-2\L g_{\m\n}&=\z\frac{\k^2}{8 \pi} \left(2 f_{\m}^{\ \a}f_{\n\a}-g_{\m\n}f_{\a\b}f^{\a\b} \right),\label{eqgravvec}\\
\p_\n(\sqdet\, f^{\n\m})&=0,\label{eqvec}
\end{align}
where $\z=\z_{1} \z_2$.
As discussed in the previous section, for a static metric in the KS form \eqref{KS}, the solution for the vector field is given in \eqref{elec}. With this at hand, one can solve the gravitational field equations \eqref{eqgravvec}. Introducing the cosmological constant modifies the left-hand side as follows:
\begin{align}
\text{(LHS)}_{tt} &= \frac{\left[\f(r)-1\right]\left[r\,\f^{\pr \pr}(r)+\f^\pr(r)-4\L r\right]}{2r},\nn\\
\text{(LHS)}_{tr} &=\frac{\f(r)\left[r\,\f^{\pr \pr}(r)+\f^\pr(r)-4\L r\right]}{2r},\nn\\
\text{(LHS)}_{rr} &=\frac{\left[\f(r)+1\right]\left[r\,\f^{\pr \pr}(r)+\f^\pr(r)-4\L r\right]}{2r}, \nn\\
\text{(LHS)}_{\t\t} &= r\, \f^\pr(r)-2\L r^2,
\end{align}
and the only nonzero component of the right-hand-side of the equations is
\begin{equation}
\RHS_{\t\t} = 2\z G q^2, 
\end{equation}
From the $\t\t$ component, the KS scalar can be solved as
\begin{equation}
\f(r) = C + \L r^2+2\z G q^2 \log(r),\qquad C = \text{constant}\label{ksvec}
\end{equation}
which also solves the other components of the field equations. In order to give a physical meaning to the integration constant $C$, we again write the metric in the BL coordinates \eqref{BL} via the transformation \eqref{trans}, which leads to the following metric function:
\begin{equation}
f(r)=1 - C - \L r^2 - 2\z G q^2 \log(r).\label{blvec}
\end{equation}
Having found the most general solution, we are now ready to investigate physically interesting possibilities.
\subsection{Solutions with the correct Newtonian potential ($\L=0$)}
In order to obtain solutions with the correct Newtonian potential, we  take $\L=0$ since only the logarithmic term is needed. In the Newtonian limit, the gravitational field in terms of the KS scalar can be obtained from Eqs. (\ref{newt})-()\ref{g}), which yields
\begin{equation}
	\vec{g} =\frac{1}{2} \vec{\nabla} \f.\label{gKS}
\end{equation}
For the KS scalar given in \eqref{ksvec} with $\L=0$, we obtain
\begin{equation}
	\vec{g} = \frac{\z G q^2}{r}\hat{r},\label{gvec}
\end{equation}
which shows that in order to preserve the attractive nature of the gravitational force, one should have $\z = -1$, i.e., either the EH term or the vector kinetic term in the action \eqref{acvec} should carry a ghost sign (see Appendix \ref{sec:app} for more details).

Note that our solution should be a charged black hole and the gravitational attraction is provided by the electric charge. Therefore, the integration constant should be a function of the mass of the black hole $C = C(M)$. In order to fix the constant, we again need to check the metric function in the BL coordinates given in \eqref{blvec} and demand that the metric reduces to the Minkowski metric [$f(r)=1$] when the mass and the charge are set to zero ($M \goesto 0$, $q \goesto 0$). This constraint can be satisfied by taking the mass term with different signs as follows:
\begin{equation}
	f^{\pm}(r)=1 \pm 8GM + 2G q^2 \log(r),\label{fpm}
\end{equation}
where for both choices $f(r)$ has a single zero, and therefore, admits one event horizon.  

\begin{figure}[!htb]
	\begin{center}
		\scalebox{0.42}{\includegraphics{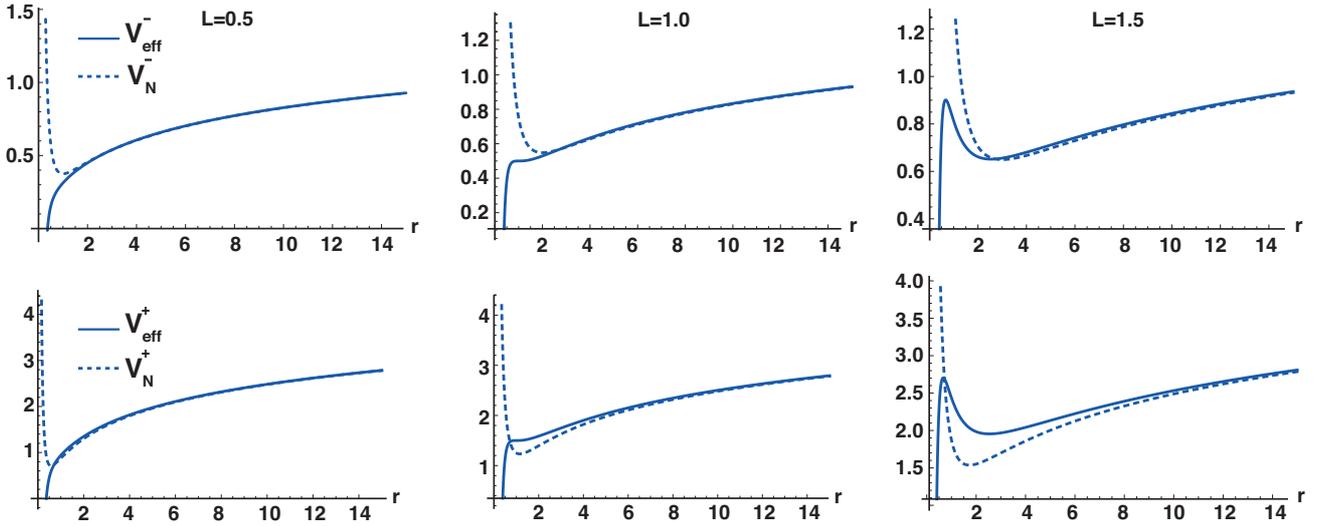}}
	\end{center}
	\caption{First row shows the effective potential $V_\eff^-$ and the Newtonian potential $V_\New^-$  for the metric function $f^-(r)$ with $GM=\frac{1}{16}$ and $Gq^2=\frac{1}{4}$. The second row shows
		the effective potential $V_\eff^+$ and the Newtonian potential $V_\New^+$ for the metric function $f^+(r)$ with $GM=\frac{1}{16}$ and $Gq^2=\frac{3}{4}$. For both cases, timelike geodesics are stable orbits when $L>L_\mi=1$.}
	\label{fig}
\end{figure}

Although we have already ensured the correct Newtonian limit, it is interesting to have a closer look at the properties of the metric as done in \cite{CarrilloGonzalez:2019gof} for the scalar case. For this purpose, we check the geodesic motion of a timelike particle described by the equation
\begin{equation}
\frac{1}{2}E^2=\frac{1}{2}\left(\frac{\dif r}{\dif t}\right)^2+V^{\pm}_\eff,\qquad V^{\pm}_\eff=\frac{1}{2}\left(\frac{L^2}{r^2}+1\right) f^{\pm}(r),\label{geod}
\end{equation}
where $E$ and $L$ are the energy and the angular momentum of the particle which are defined through the timelike and the angular  Killing vectors $\xi^{\m}_{(t,\t)}$ as follows:
\begin{equation}
	E = -g_{\m\n}\,\xi^{\m}_{(t)}u^\n,\qquad L=g_{\m\n}\,\xi^{\m}_{(\t)}u^\n,
\end{equation}
where $u^\m$ is the velocity the particle. The Newtonian limit of the effective potential $V^{\pm}_\eff$ can be obtained by neglecting the $L^2G$ terms as follows\footnote{For a general analysis, one should write the logarithmic term in both $V_\eff$ and $V_\New$ by introducing a length scale as $\log(\frac{r}{r_0})$. We set $r_0=1$ for simplicity.}:
\begin{equation}
	V^{\pm}_\New = \frac{1}{2} \pm 4GM + \frac{L^2}{2r^2}+Gq^2 \log(r).
\end{equation}
The metric function obtained by coupling a free scalar \eqref{fsca} and the ones we obtained by coupling a gauge vector \eqref{fpm} have the same functional form [$f(r)=A + B \log(r)$, $A, B$: constant]. Therefore, all the physically important properties of the solution that is discussed in \cite{CarrilloGonzalez:2019gof} are also valid for our solutions.  They can be summarized as follows:
\begin{enumerate}
	\item The Newtonian potential $V_\New$ has an infinite barrier at short distances and matches with the effective potential $V_\eff$ at long distances.
	\item A timelike particle cannot escape to infinity due to the logarithmic divergence of the potential as $r \goesto \infty$.
	\item The effective potential $V_\eff$ develops a local maximum ($V^\ma_\eff$) and a local minimum ($V^\mi_\eff$) when the angular momentum of the particle $L$ is larger than a certain value ($L_{min}$). A timelike particle moves along a stable orbit provided that $V^\mi_\eff \leq E<V^\ma_\eff$. On the other hand, the Newtonian potential $V_\New$ always admit stable orbits.
	\item When $E=V^\mi_\eff$ and $L>L_\mi$, timelike geodesics form circular orbits, i.e., one has $\frac{\dif r}{\dif t}=0$ in \eqref{geod}. 
	\item Since the central potential is not that of an inverse-square central force [$V(r)=-\frac{k}{r}$, $k$: constant] or a radial harmonic oscillator [$V(r)=\frac{1 }{2} k r^2$, $k$: constant], Bertrand's theorem assures that there will be precession for orbits with $E>V^\mi_\eff$.
\end{enumerate}
In Fig. \ref{fig}, we show that properties 1-4 hold for the metric functions $f^{\pm}(r)$ given in \eqref{fpm} by tuning the parameters such that $L_\mi=1$. Having shown that the qualitative properties of the metric is the same, we refer the reader to \cite{CarrilloGonzalez:2019gof} where the authors present timelike geodesics, and also, show that more precession is observed in the relativistic orbits when compared to the Newtonian orbits.

The KS scalars corresponding to the metric functions \eqref{fpm} are given by
\begin{equation}
	\f^{\pm}(r)=\mp 8GM - 2Gq^2 \log(r),\label{KS3}
\end{equation}
and lead to the following single copy gauge field:
\begin{equation}
	A_\m\dif x^\m =\f k_\m\dif x^\m=  (\pm 8GM + 2Gq^2\log r) \,(\dif t + \dif r).
\end{equation}
The constant factor does not play a role and this is just the Coulomb solution \eqref{coulomb} with the identification $2Gq^2 \goesto Q$, i.e., the electric charge in the gravity side $q$ yields a positively charged point particle in the gauge theory. This is a remarkable difference compared to the higher dimensional cases.\footnote{Various aspects of the charged black holes solutions in the context of the KS double copy are discussed in \cite{Carrillo-Gonzalez:2017iyj} and the source terms for $d=4$ are given in \cite{Alkac:2021bav} .} In dimensions higher than three ($d \geq 4$), the static solution of the Einstein-Maxwell theory, the Reissner-Nordstr\"{o}m black hole, has the following KS scalar:
\begin{equation}
\f(r) = \frac{2GM}{r^{d-3}}-\frac{G q^2}{r^{2(d-3)}},\label{KShigh}
\end{equation}
where $M$ and $q$ are the mass and the electric charge of the black hole respectively. The gauge field in the gravity side and the corresponding field strength tensor are given by
\begin{align}
a_\m \dif x^\m &= -\frac{q}{r^{(d-3)}} \dif t,\label{Coulhigh}\\
\frac{1}{2}\,f_{\m\n}\, \dif x^\m \wed \dif x^\n &= (d-3)\frac{\,q}{r^{d-2}}\, \dif r \wed \dif t.
\end{align}
The gauge theory source for the solution is as follows:
\begin{equation}
J^\m = \r\, \d^{\m}_{\ 0}, \qquad \qquad \r = 2G M \d^{d-1}(\vec{r}) - \frac{2(d-3)^2Gq^2}{r^{2(d-2)}},\label{Jhigh}
\end{equation}
where we see that the mass $M$ of the black hole shows itself as the charge of a point particle and the electric charge $q$ results in a nonlocalized charge distribution which vanishes as $r \goesto \infty$. Obtaining the Coulomb solution in $d=3$ is a very peculiar property, which is possible thanks to the fact that the existence of the electric charge in the gravity side changes the KS scalar \eqref{KS3} such that the modification has the same functional form [$\log(r)$] with the Coulomb solution in $d=3$. In higher dimensions, as can be seen in \eqref{KShigh}, the mass term carries the functional form of the Coulomb solution given in \eqref{Coulhigh}, and therefore, yields a point charge in the gauge theory. The modification due to the electric charge has a different functional form and produces a nonlocalized charge density as described in \eqref{Jhigh}.

\subsection{The charged Ba\~nados-Teitelboim-Zanelli (BTZ) black hole ($\L<0$)}
We have shown that the Coulomb solution can be obtained as a gauge theory single copy by considering Einstein-Maxwell theory where either the EH term or the vector kinetic term carries a ghost sign. One can introduce the cosmological constant $\L$ such that solutions with the correct Newtonian potential are recovered when $\L=0$. Instead, we will study the charged BTZ black hole whose metric function reads \cite{Banados:1992wn}
\begin{equation}
	f(r)=-8 G M + \frac{r^2}{\ell^2} - 2 G q^2 \log(r).\label{blbtz}
\end{equation}
Comparing this with the most general solution \eqref{blvec} gives that the parameters should be chosen as follows:
\begin{equation}
	C=1+8GM,\qquad \L=-\frac{1}{\ell^2},\qquad \z=1, 
\end{equation}
where the last one equation shows that no ghost field is needed to obtain the solution. 

Gravitational field equations with the cosmological constant \eqref{eqgravvec} can be mapped to Maxwell's equations by again checking the $\m0$ component of the trace-reversed equations and using \eqref{delF}, which yield
\begin{equation}
\p_\n F^{\n\m}=g\left[J^\m_{(\L=0)}+\bar{J}^\m\right].
\end{equation}
Here, $J^\m_{(\L=0)}$ is the source in the absence of the cosmological constant, whose general form is given in \eqref{Jdef}.  $\bar{J}^\m$ represents the effect of the cosmological constant on the source and takes the following form:
\begin{equation}
	\bar{J}^\m = \r_c \bar{v}^\m, \qquad \r_c=\frac{4\L}{g}, \qquad \bar{v}^\m=\left(1,0,0\right),
\end{equation}
which is a constant charge density filling all space. The KS scalar corresponding to the metric function \eqref{blbtz} is
\begin{equation}
\f(r) = 1+8GM + \L r^2+2 G q^2 \log(r),
\end{equation}
with the single copy gauge field
\begin{equation}
	A_\m\dif x^\m =\f k_\m\dif x^\m= - \left[1+8GM + \L r^2+2 G q^2 \log(r)\right] \,(\dif t + \dif r).
\end{equation}
The field strength tensor reads
\begin{equation}
	\frac{1}{2}\,F_{\m\n}\, \dif x^\m \wed \dif x^\n = \left[\frac{Q}{r}-\L r\right]\, \dif r \wed \dif t,
\end{equation}
where we have made the replacement $-2Gq^2 \goesto Q$. We see that the gauge theory single copy of the charged BTZ black hole is the
Coulomb solution ($Q<0$) modified by a term which describes an electric field linearly increasing with the radial coordinate $r$ (since $\L<0$) and the source is a point charge located in a medium of constant charge density as follows:
\begin{equation}
	J^\m \p_\m = \left[Q\, \d^2(\vec{r})+\frac{4\L}{g}\right]\, \p_t. 
\end{equation}
It is important to note that this is the usual behavior of the Schwarzschild-AdS black hole in higher dimensions when written around a flat background metric \cite{Alkac:2021bav}.

 As we see, the Coulomb solution ($Q<0$) modified by the cosmological constant can be obtained from the well-known charged BTZ black hole without any need for introducing a ghost. From \eqref{gKS}, one can calculate the gravitational field in the Newtonian limit as
\begin{equation}
	\vec{g} = \left[\frac{\L r}{2}+\frac{ G q^2}{r}\right]\hat{r},
\end{equation}
which shows that a negative cosmological constant ($\L<0$) is needed for an attractive force, which is possible when $r>\sq{\frac{2Gq}{-\L}}$. The geodesics of the charged BTZ black hole exhibit a very rich structure and the details can be found in \cite{Soroushfar:2015dfz}.
\section{Conclusions}\label{sec:conc}	
In this paper, we have studied the KS double copy of the Coulomb solution	in 3d, which is an important consistency check for the classical double copy due to the lack of degrees of freedom and a Newtonian limit in GR. The double copy solution should have the correct Newtonian limit, and in 3d, this can only be achieved by matter coupling. In \cite{CarrilloGonzalez:2019gof}, the solution was constructed by coupling to a scalar but has some undesired features. It is a hairy black hole which requires that either the EH term or the scalar kinetic term carry a ghost sign, and the scalar field does not vanish at infinity. Making use of the on-shell duality of a free scalar and a gauge vector, we have shown that a solution with the correct Newtonian limit can also be obtained as a solution of Einstein-Maxwell theory such that the single copy is again the Coulomb solution. While at least one ghost sign is still needed, the electric field in the gravity side vanishes at infinity, which is an improvement compared to the scalar case. 

When a negative cosmological constant is introduced, the charged BTZ black hole is a solution without any need for a ghost field, and we have shown that the single copy gauge field is the Coulomb solution ($Q<0$) modified by a term describing an electric field whose magnitude linearly increases with the distance to the point charge. The source is a point particle sitting in a medium of constant charge density, which is the usual effect of the cosmological constant. At the expense of this modification, this remarkably establishes a connection to the well-known black hole solutions in 3d gravity, which, we believe, shows the potential of 3d KS double copy to have many other interesting features.  We hope to report more on this in the future.

\section*{Acknowledgments}
	M. K. G. is supported by T\"{U}B\.{I}TAK Grant No 118F091. G.A. thanks Mehmet \"{O}zkan for helpful discussions regarding the on-shell duality of $p$-form gauge fields. We thank Merve Demirta\c{s} Alka\c{c} for her help in creating high-resolution figures.

\appendix
\section{Scalar-Vector Duality as Described in \cite{Bueno:2021krl}}\label{sec:app}

Note that in our formulation of the scalar-vector duality in Sec. \ref{sec:dual}, we have shown that free field equations in flat spacetime are equivalent and a scalar field which is linear in the azimutal angle ($\vf = c\, \t$, $c=\cons$) implies the Coulomb solution for the gauge vector. This can be generalized to curved spacetime as long as the metric is in the KS form \eqref{KS}. However; our analysis is limited to the matter equations and the gravity equations have to be checked independently. The action \eqref{dualact} that we use to establish the duality leads to a free scalar action with a ghost sign \eqref{freesca} and a free Maxwell action with a nonghost sign \eqref{freevec},  which is enough to show the equivalence of the matter equations. When coupling to the gravitation is considered, in order to obtain a black hole solution with the same physical properties, one needs to have a ghost scalar and a ghost vector (we assume $\z_1=1$, i.e., the EH term has the ``right sign,'' for simplicity.)

Indeed, as shown in \cite{Bueno:2021krl} recently,  the duality can also be formulated such that the gravitational action with the correct sign for the matter coupling is obtained directly. The authors consider theories described by Lagrangians of the form $\cL(g^{\m\n},R_{\m\n},\p_\m \vf)$ and study solutions in the following form
\begin{equation}
	\dif s^2 = -f(r) \dif t^2 + \frac{\dif r^2}{f(r)}+ r^2 \dif \t^2, \qquad \vf = c\, \t,\qquad c=\cons,
\end{equation}
which includes the solution that we studied in Sec. \ref{sec:sca} with the metric written in the BL coordinates. It is possible to find theories such that the scalar equations is automatically satisfied, the equation for the metric function gets a nontrivial modification and can be solved analytically. They also show that the same solution can be obtained from a dual Lagrangian of the form $\cL_{\dual}(g^{\m\n},R_{\m\n},f_{\m\n})$. When the matter fields are related as\footnote{We use a different normalization than \cite{Bueno:2021krl} to obtain the vector kinetic term with coefficient $\frac{1}{8\pi}$, as we have used throughout the text.}
\begin{equation}
f_{\m\n} = -\sq{2\pi}\, \ep_{\m\n\r} \frac{\p \cL}{\p(\p_\r \vf)}\label{newdual}
\end{equation}
the dual Lagrangian is given by
\begin{equation}
\cL_\dual = \cL - \frac{1}{2\sq{2\pi}}\ep^{\m\n\r} f_{\m\n} \p_\r \vf,
\end{equation}
where the original Lagrangian $\cL$ should also be written in terms of $f_{\m\n}$ by using the relation \eqref{newdual}. In \cite{Bueno:2021krl}, nonminimal matter couplings are used and regular, electrically charged black hole solutions in three dimensions are obtained. In this work, we only have the kinetic terms, which are the simplest possible matter couplings. Starting from the Lagrangian
\begin{equation}
\cL=\frac{1}{\kappa^{2}}R+\frac{1}{2}\left(\p \vf\right)^2, 
\end{equation}
leads to the following dual Lagrangian
\begin{equation}
\cL_\dual=\frac{1}{\k^2}R+\frac{1}{8 \pi}f_{\m\n}f^{\m\n}.
\end{equation}
Both Lagrangians $\cL$ and $\cL_\dual$ support the solution with the correct Newtonian potential as long as the electric charge $q$ and the constant $c$ are related as 
\begin{equation}
	c=\frac{q}{\sq{2\pi}},
\end{equation}
which is a consequence of \eqref{newdual}. Using the relation between the mass parameter $M$ in the scalar case  with the constant $c$ \eqref{scaparam}, this implies 
\begin{equation}
	M = q^2,
\end{equation}
which shows the relation between the coefficient of the logarithmic terms in the metric functions \eqref{fsca} and \eqref{fpm}.
%}

\end{document}